\documentclass[aps,pra,twocolumn,superscriptaddress]{revtex4-1}

\usepackage{amsmath}
\usepackage{cases}
\usepackage{graphicx}
\usepackage{mathtools} 
\usepackage[T1]{fontenc} 
\usepackage[utf8]{inputenc} 
\usepackage{amsfonts} 
\usepackage{bm}

\usepackage{enumitem}
\usepackage{multirow}
\usepackage{tabularx}
\usepackage{longtable}
\usepackage{float}

\usepackage{hyperref}
\usepackage[capitalise,nameinlink]{cleveref}
\usepackage[usenames,dvipsnames]{xcolor}
\hypersetup{
    colorlinks=true,       
    linkcolor=Maroon,          
    citecolor=OliveGreen,        
    filecolor=magenta,      
    urlcolor=Blue           
}
\crefname{ansatz}{Ansatz.}{Ansatzes.}

\def\ket#1{\left|#1\right\rangle}
\def\bra#1{\langle#1|}

\newcommand{\QuArC}{%
    \affiliation{%
        Microsoft Quantum,
        Redmond WA 98052, USA.}}
\newcommand{\PNNL}{%
    \affiliation{%
        William R. Wiley Environmental Molecular Sciences Laboratory, Battelle, 
        Pacific Northwest National Laboratory, K8-91, P.O. Box 999, Richland WA 99352, USA}}

\begin{document}


\title{Quantum simulations of excited states with active-space downfolded Hamiltonians}

\author{Nicholas P. Bauman} \PNNL
\author{Guang Hao Low}
\QuArC
\author{Karol Kowalski} 
\email{karol.kowalski@pnnl.gov}
\PNNL


\date{\today}

%
%

\begin{abstract}
Many-body techniques based on the double unitary coupled cluster ansatz (DUCC) can be 
used to downfold electronic Hamiltonians into low-dimensional active spaces. It can be 
shown that the resulting dimensionality reduced Hamiltonians are amenable for quantum 
computing. Recent studies performed for several benchmark systems using quantum 
phase estimation (QPE) algorithms demonstrated that these formulations can recover a significant 
portion of ground-state dynamical correlation effects that stem from the electron 
excitations outside of the active space. These results have also been confirmed in 
studies of ground-state potential energy surfaces using quantum simulators. In this letter, 
we study the effectiveness of the DUCC formalism in describing excited states. We also 
emphasize the role of the QPE formalism and its stochastic nature in discovering/identifying 
excited states or excited-state processes in situations when the knowledge about the 
true configurational structure of a sought after excited state is limited or postulated (due to 
the specific physics driving excited-state processes of interest). In this context, we 
can view the QPE algorithm as an engine for verifying various hypotheses for excited-state 
processes and providing statistically meaningful results that correspond to the electronic 
state(s) with the largest overlap with a postulated configurational structure. We illustrate 
these ideas on examples of strongly correlated molecular systems, characterized by 
small energy gaps and high density of quasi-degenerate states around the Fermi level. 
\end{abstract}


\maketitle


\section{Introduction}
There is significant interest in applying quantum computing techniques to 
describe and simulate chemical systems and processes~\cite{ortiz_gubernatis,aspuru2005simulated,lanyon2010towards,seeley2012bravyi,peruzzo2014variational,wecker2015progress,Babbush2016exponentially,mcclean2016theory,havlivcek2017operator,fontalvo2017strategies,openfermion,PhysRevA.95.020501,reiher2017elucidating,setia2018bravyi,low_depth_Chan,motta2018low,mayhall2019}.
The approach brings hope to addressing the exponential barriers limiting the applicability 
of exact diagonalization procedures (or full configuration interaction methods), and also to
provide access to complicated multi-configurational electronic states, which often can not 
be identified or described by vast classes of approximate methods used in routine simulations. 
Of special interest is the application to strongly correlated molecular systems 
characterized by small energy gaps between occupied and unoccupied orbitals where multiple 
electronic states that are defined by complex collective phenomena exist, usually involving 
higher than single excitations in the corresponding wavefunction expansions.

Even though impressive progress has been achieved in the development of wavefunction-based 
excited-state approaches such as complete-active-space perturbation theory (CASPT)~\cite{andersson1990second,andersson1992second},
multi-reference NEVPT~\cite{angeli2001n,angeli2001introduction},
configuration interaction~\cite{werner1988efficient},
equation-of-motion coupled cluster (EOMCC)~\cite{bartlett07_291,bartlett89_57,bartlett93_414,stanton93_7029, piecuch1999eomxcc,krylov2001size,hirata2004higher},
multi-reference coupled cluster (MRCC)~\cite{dm1975,kaldor1991fock,rittby1991multireference,meissner1998fock,musial2008multireference,jezmonk,meissner1,pylypov1,pylypov2,piepal1992,mahapatra1,mahapatra2,evangelista1,bwpittner1,hanauer2011pilot,AK58,koehn2013state,mrcclyakh},
and the density matrix renormalization group (DMRG)~\cite{white1992density,schollwock2005density,dmrg2,chan2011density},
problems with the description of complicated states dominated by high-rank excitations still exist. 
For example, in order to capture these states with EOMCC formalisms one needs to include 
higher-than-double excitations~\cite{watts1999equation}. 
For doubly excited states, the "minimum" level of theory to tackle these states is EOMCC with 
singles, doubles, and triples (EOMCCSDT)~\cite{kkppeom,kucharski2001coupled} although in several cases it may not provide 
a quantitative level of  accuracy~\cite{bhaskaran2012note}. These problems commonly occur even for small 
molecular systems and it is reasonable to expect that they may intensify for larger systems and 
strongly correlated systems (transition metal oxides, metal clusters, actinides) with a high 
density of states located around the Fermi level. 
Recently, a significant progress in addressing these problems  has been achieved by integrating stochastic  configuration interaction (CI) Quantum Monte Carlo (QMC) framework 
\cite{booth2009fermion,cleland2010communications} with deterministic EOMCC formulations.\cite{deustua2019accurate}
Another interesting aspect of modeling these complex excited 
states is the attainability of these states in situations where their initial configurational 
structure cannot be easily obtained to commence convergent iterative procedures. 

Progress in the development of quantum algorithms may 
provide alternative solutions to these problems. The recent application of variational quantum 
eigensolvers (VQE)~\cite{peruzzo2014variational,mcclean2016theory,openfermion,romero2018strategies,PhysRevA.95.020501,kandala2018extending,PhysRevX.8.011021} and quantum phase estimation (QPE)~\cite{luis1996optimum,cleve1998quantum,berry2007efficient,childs2010relationship,wecker2015progress,Kimmel2015robust,Wiebe2016bayesian,haner2016high,poulin2018quantum} 
to excited states~\cite{PhysRevX.8.011021,low2019q}
demonstrate 
that excited states can be effectively simulated on the quantum computers while at the same time bypassing
the problems of conventional computing and approximate formulations.

In this paper, we present an excited state extension of recently developed techniques for active-space downfolding of the electronic Hamiltonian based 
on the double unitary coupled cluster (DUCC) transformation.
This is combined with QPE simulations available in the Microsoft Quantum Development Kit (QDK)~\cite{low2019q,Low2016qubitization} to illustrate the excited-state version of DUCC formalism on the examples of H$_2$ at equilibrium and stretched
bond lengths and two H$_4$ models: (1) trapezoidal H$_4$, which is a popular benchmark system for studying 
quasi-degenerate states~\cite{jankowski1980applicability} and (2) a linear H$_4$ molecule, which was intensively studied in the context 
of singlet fission processes~\cite{minami2011diradical}. For each of these system, we perform QPE simulations to characterize the 
structure of excited states using DUCC effective Hamiltonians. Moreover, we carry out a number of QPE 
simulations to investigate the role and effect of different initial guesses which are based on limited knowledge 
about the excited state of interest.

\section{Theory of DUCC Downfolded Hamiltonians}

In Ref.~\cite{bauman2019downfolding}, we introduced the unitary extension of the sub-system embedding sub-algebra CC approach (SES-CC)~\cite{safkk} which utilizes the double unitary CC expansion
\begin{equation}
	\ket{\Psi}=e^{\sigma_{\rm ext}} e^{\sigma_{\rm int}}\ket{\Phi}.
\label[ansatz]{ducc1}
\end{equation}
The character of the expansion (\ref{ducc1})  is similar to the  expansion  discussed in the 
single-reference formulation of the active-space coupled cluster formalism,\cite{active3, activerev}  
(see also Refs. \cite{ active1, active2})  which also utilizes the decomposition of the cluster operator into internal and external parts.

In analogy to Ref.~\cite{bauman2019downfolding},  $\sigma_{\rm int}$ and $\sigma_{\rm ext}$ are the anti-Hermitian operators 
($\sigma_{\rm int}^{\dagger}=-\sigma_{\rm int}$ and $\sigma_{\rm ext}^{\dagger}=-\sigma_{\rm ext}$)
defined by excitations/de-excitations within and outside of active space, respectively. To be more 
precise, the amplitudes defining the $\sigma_{\rm ext}$ operator must carry at least one inactive spin-orbital index. 
Using~\cref{ducc1} in Schr\"odinger's equation, one obtains equations for cluster amplitudes and the 
corresponding energy
\begin{align}
        Qe^{-\sigma_{\rm int}}e^{-\sigma_{\rm ext}} H e^{\sigma_{\rm ext}}e^{\sigma_{\rm int}} \ket{\Phi} &= 0,
\label{uccd2eq} \\
        \bra{\Phi}e^{-\sigma_{\rm int}}e^{-\sigma_{\rm ext}} H e^{\sigma_{\rm ext}}e^{\sigma_{\rm int}} \ket{\Phi} &= E,
\label{uccd2ene}
\end{align}
where $Q$ is a projection operator on the space spanned by determinants orthogonal to the reference function $\ket{\Phi}$.
In these and subsequent equations, we consider the case of the exact limit ($\sigma_{\rm int}$ and 
$\sigma_{\rm ext}$ include all possible excitations). 
In Ref.~\cite{bauman2019downfolding}, we showed that when $\sigma_{\rm int}$ contains all possible excitations/de-excitations within the 
complete active space, the energy of the system~\cref{uccd2eq} can be obtained by diagonalizing
the DUCC effective Hamiltonian
\begin{equation}
        \bar{H}_{\rm ext}^{\rm eff(DUCC)} e^{\sigma_{\rm int}} \ket{\Phi} = E e^{\sigma_{\rm int}}\ket{\Phi},
\label{duccstep2}
\end{equation}
where
\begin{equation}
        \bar{H}_{\rm ext}^{\rm eff(DUCC)} = (P+Q_{\rm int}) \bar{H}_{\rm ext}^{\rm DUCC} (P+Q_{\rm int})
\label{equivducc}
\end{equation}
and 
\begin{equation}
        \bar{H}_{\rm ext}^{\rm DUCC} =e^{-\sigma_{\rm ext}}H e^{\sigma_{\rm ext}}.
\label{duccexth}
\end{equation}
In the above eigenvalue  problem, the   $e^{\sigma_{\rm int}}\ket{\Phi}$   expansion defines  a corresponding eigenvector and  $P$ and $Q_{\rm int}$ are  projection operators onto the reference function $|\Phi\rangle$ and excited determinants in the active space orthogonal to $\ket{\Phi}$, respectively.  

To show this property  it is sufficient  to introduce the resolution of identity $e^{\sigma_{\rm 
int}}e^{-\sigma_{\rm int}}$ to the left of the $\bar{H}_{\rm ext}^{\rm DUCC} $ operator 
in 
\begin{equation}
	(P+Q_{\rm int}) \bar{H}_{\rm ext}^{\rm DUCC} e^{\sigma_{\rm int}} |\Phi\rangle = E
    (P+Q_{\rm int}) e^{\sigma_{\rm int}}|\Phi\rangle\;,
\label{duccstep1}
\end{equation}
where we explicitly used the property of the $e^{\sigma_{\rm int}}|\Phi\rangle$ expansion
\begin{equation}
(P+Q_{\rm int}) e^{\sigma_{\rm int}}|\Phi\rangle =
e^{\sigma_{\rm int}}|\Phi\rangle \;,
\label{pqipro}
\end{equation}
and to notice that $e^{-\sigma_{\rm int}}\bar{H}_{\rm ext}^{\rm 
DUCC} e^{\sigma_{\rm int}}=e^{-\sigma_{\rm int}}e^{-\sigma_{\rm ext}} H e^{\sigma_{\rm 
ext}}e^{\sigma_{\rm int}}$. 
Next, using matrix representation of the $\sigma_{\rm int}$ operator in the CAS space denoted 
as $\bm{\sigma}_{\rm int}$ this equation can be re-written as 
\begin{equation}
	[e^{\boldmath{\sigma}_{\rm int}}] [\boldmath{y}] = 0 \;,
\label{llineq2}
\end{equation}
where the first component of the $[\bm{y}]$ vector is equivalent to 
$\langle\Phi|e^{-\sigma_{\rm int}}e^{-\sigma_{\rm ext}} H 
e^{\sigma_{\rm ext}}e^{\sigma_{\rm int}} |\Phi\rangle-E$
while the remaining components correspond to projections of 
$e^{-\sigma_{\rm int}}e^{-\sigma_{\rm ext}} H 
e^{\sigma_{\rm ext}}e^{\sigma_{\rm int}} |\Phi\rangle$
onto excited configurations belonging to $Q_{\rm int}$. 
The $[e^{\boldmath{\sigma}_{\rm int}}]$ matrix is also  non-singular, which is a consequence of the formula 
\begin{equation}
	{\rm det}(e^{\bm{\sigma}_{\rm int}})=e^{{\rm Tr}(\bm{\sigma}_{\rm int})} =1
\label{det1}
\end{equation}
and the anti-Hermitian character of the $\bm{\sigma}_{\rm int}$
matrix, i.e., ${\rm Tr}(\bm{\sigma}_{\rm int})=0$
(where real character of $\sigma_{\rm int}$ cluster amplitudes is assumed).
Given the non-singular character 
of the $[e^{\boldmath{\sigma}_{\rm int}}]$ matrix 
(see also  Ref.~\cite{bauman2019downfolding}), this proves the equivalence of these two representations. 

The proof of the above property is not limited to the ground state and can be extended to any electronic state described by~\cref{ducc1} and~\cref{uccd2eq,uccd2ene}. 
Assuming that this ansatz can describe excited states, the DUCC effective Hamiltonian formalism can be used in the context of excited-state
simulations.  We will denote general DUCC  solution corresponding to the $K$-th state as 
\begin{equation}
\ket{\Psi(K)} = e^{\sigma_{\rm ext}(K)} e^{\sigma_{\rm int}(K)} \ket{\Phi},
\label[ansatz]{psik}
\end{equation}
where the $K$-th state energy can be obtained from diagonalizing the state-specific effective Hamiltonian
\begin{equation}
        \bar{H}_{\rm ext}^{\rm eff(DUCC)} (K) e^{\sigma_{\rm int}(K)} \ket{\Phi} = E_K e^{\sigma_{\rm int}(K)}\ket{\Phi},
\label{duccstep2}
\end{equation}
where
\begin{equation}
        \bar{H}_{\rm ext}^{\rm eff(DUCC)}(K) = (P+Q_{\rm int}) \bar{H}_{\rm ext}^{\rm DUCC} (K)(P+Q_{\rm int}),
\label{equivducc}
\end{equation}
and 
\begin{equation}
        \bar{H}_{\rm ext}^{\rm DUCC}(K) =e^{-\sigma_{\rm ext}(K)}H e^{\sigma_{\rm ext}(K)}.
\label{duccexth}
\end{equation}
Similar to the ground-state effective/downfolded Hamiltonians, the operators
$ \bar{H}_{\rm ext}^{\rm eff(DUCC)}(K)$ are Hermitian and therefore amenable to real-time simulation on a quantum computer.
In analogy to the ground-state representation of DUCC, we will assume that 
\begin{equation}
\sigma_{\rm int}(K)^{\dagger} = -\sigma_{\rm int}(K)  \;\;\;\; \sigma_{\rm ext}(K)^{\dagger} = -\sigma_{\rm ext}(K) 
\label{sigmak1}
\end{equation}
and 
\begin{eqnarray}
\sigma_{\rm int}(K)&=& S_{\rm int}(K)-S_{\rm int}(K)^{\dagger}  , \label{sigmak2} \\
\sigma_{\rm ext}(K)&=& S_{\rm ext}(K)-S_{\rm ext}(K)^{\dagger} , \label{sigmak3} 
\end{eqnarray}
where $S_{\rm int}(K)$ and $S_{\rm ext}(K)$ are CC-type cluster operators producing excitations within and outside the 
active space, respectively, when acting on the reference function $\ket{\Phi}$.

If the exact form of the operator $\sigma_{\rm ext}(K)$ (or $S_{\rm ext}(K)$) is known, the 
effective Hamiltonian $\bar{H}_{\rm ext}^{\rm eff(DUCC)}(K)$ can be diagonalized to find corresponding 
exact energy $E_K$. In practice, in likeness to the ground-state formulation, we obtain an
approximate model $\sigma_{\rm ext}(K)$ operator by way of calculations with excited-state CC models. 
Additionally, we previously employed in ground-state DUCC~\cite{bauman2019downfolding}
the many-body form of the effective Hamiltonian $\bar{H}_{\rm ext}^{\rm eff(DUCC)}$ driven by the perturbative analysis of the 
ground-state energy expansion. 
However, the same arguments cannot be invoked in the context of the excited-state variant of $\bar{H}_{\rm ext}^{\rm DUCC}(K)$.
Instead, in the analysis of the excited states we use a basic expansion for $\bar{H}_{\rm ext}^{\rm DUCC}(K)$ given by 
an expression involving a single commutator:
\begin{eqnarray}
        \bar{H}_{\rm ext}^{\rm DUCC}(K) &=& e^{-\sigma_{\rm ext}(K)}H e^{\sigma_{\rm ext}(K)} \nonumber \\
        &\simeq& H+[H,\sigma_{\rm ext}(K)].
\label{duccexth2}
\end{eqnarray}

In this paper, we explore a simple strategy based on the utilization of the excited-state wavefunction in the equation-of-motion CC parametrization 
$\ket{\Psi_K^{\rm EOMCC}}$, 
\begin{equation}
\ket{\Psi_K^{\rm EOMCC}} = R_K e^T \ket{\Phi},
\label{eomcc} 
\end{equation}
as a reference to  extract the relevant information about $\sigma_{\rm ext}(K)$. 
In the above equation, the cluster operator $T$ satisfies the CC equations and the excitation operator $R_K$ (corresponding to $K$-th excited state) 
is obtained through diagonalization of the similarity transformed Hamiltonian $\bar{H}=e^{-T}He^T$. 
Since~\cref{psik} represents a normalized state, in order  to compare with the corresponding EOMCC expansion in the exact wavefunction limit, one needs to use a normalized form $\ket{\tilde{\Psi}_K^{\rm EOMCC}}$ of~\cref{eomcc} 
\begin{equation}
\ket{\tilde{\Psi}_K^{\rm EOMCC}}=N_K R_K e^T \ket{\Phi} = e^{\sigma_{\rm ext}(K)} e^{\sigma_{\rm int}(K)} \ket{\Phi} ,
\label{norm1}
\end{equation}
where 
\begin{equation}
N_K=\frac{1}{\sqrt{\langle\Psi_K^{\rm EOMCC}\ket{\Psi_K^{\rm EOMCC}}}}.
\label{norm2}
\end{equation}
In the simplest approximate variants we use the EOMCCSD approximations (EOMCC with singles and doubles~\cite{bartlett89_57}) to extract 
singly- ($S_{\rm ext,1}(K)$) and doubly-excited  ($S_{\rm ext,2}(K)$)  components of the $S_{\rm ext}(K)$ operator
\begin{align}
S_{\rm ext,1}(K) \ket{\Phi} &\simeq   Q_{\rm ext,1} \ket{\tilde{\Psi}_K^{\rm EOMCC}}, \label{s1ext} \\
S_{\rm ext,2}(K) \ket{\Phi} &\simeq   Q_{\rm ext,2} \ket{\tilde{\Psi}_K^{\rm EOMCC}}, \label{s2ext} 
\end{align}
where $Q_{\rm ext,1}$ and $Q_{\rm ext,2}$ are projections operator on subspaces of singly and doubly excited external excitations, respectively. 
In~\cref{s1ext} and~\cref{s2ext}, we approximate $\ket{\tilde{\Psi}_K^{\rm EOMCC}}\approx\ket{\Psi_K^{\rm EOMCCSD}(A)}$, which is defined in the following way: 
\begin{eqnarray}
&&\ket{\Psi_K^{\rm EOMCCSD}(A)} = \nonumber\\
&&\;\;\;(P+Q_1+Q_2) (R_{K,0}+R_{K,1}+R_{K,2}) e^{T_1+T_2}\ket{\Phi}, \quad
\label{eomccsd_a}
\end{eqnarray}
where $Q_1$ and $Q_2$ operators are projection operators on spaces of singly and doubly excited configurations, 
$R_{K,i}$ ($i=0,1,2$) represent EOMCC excitation operators producing $i$-tuply excited configurations when acting onto reference functions, and
$T_1$ and $T_2$ are the singly and doubly excited cluster operators. In our approximation, referred to as the  DUCC-ex($K$) approach, 
we use $T_1$ and $T_2$ from standard CCSD calculations, and $R_{K,i}$ ($i=0,1,2$) from standard EOMCCSD diagonalization procedure.
Consequently, $S_{\rm ext,1}(K)$ and $S_{\rm ext,2}(K)$ take the following form 
\begin{align}
S_{\rm ext,1}(K) \ket{\Phi} &\simeq   N_K(A) Q_{\rm ext,1} (R_{K,0}T_1+R_{K,1}) \ket{\Phi}, \label{s1exp} \\
S_{\rm ext,2}(K) \ket{\Phi} &\simeq   N_K(A) Q_{\rm ext,2} (R_{K,0}(T_2+\frac{1}{2} T_1^2) \nonumber \\
&\quad     + R_{K,1}T_1 + R_{K,2}) \ket{\Phi}, \label{s2exp} 
\end{align}
where 
\begin{equation}
N_K(A)=\frac{1}{\sqrt{\langle\Psi_K^{\rm EOMCC}(A)\ket{\Psi_K^{\rm EOMCC}(A)}}}.
\label{norm2}
\end{equation}
As in the ground-state case, the DUCC-ex($K$) approximation is defined by the length of the commutator
expansion and the source of the $\sigma_{\rm ext}(K)$ amplitudes. Since the EOMCCSD approximation is 
used to define $\sigma_{\rm ext}(K)$, one should expect that this scheme to work in cases where 
EOMCCSD approach delivers reliable results and active-space used in the state-selective DUCC-ex($K$)
formalism capable to capture main configurations corresponding to $K$-th state. 


\section{Trial Wavefunction}
\begin{figure}
\includegraphics[clip,angle=0, width=0.45\textwidth]{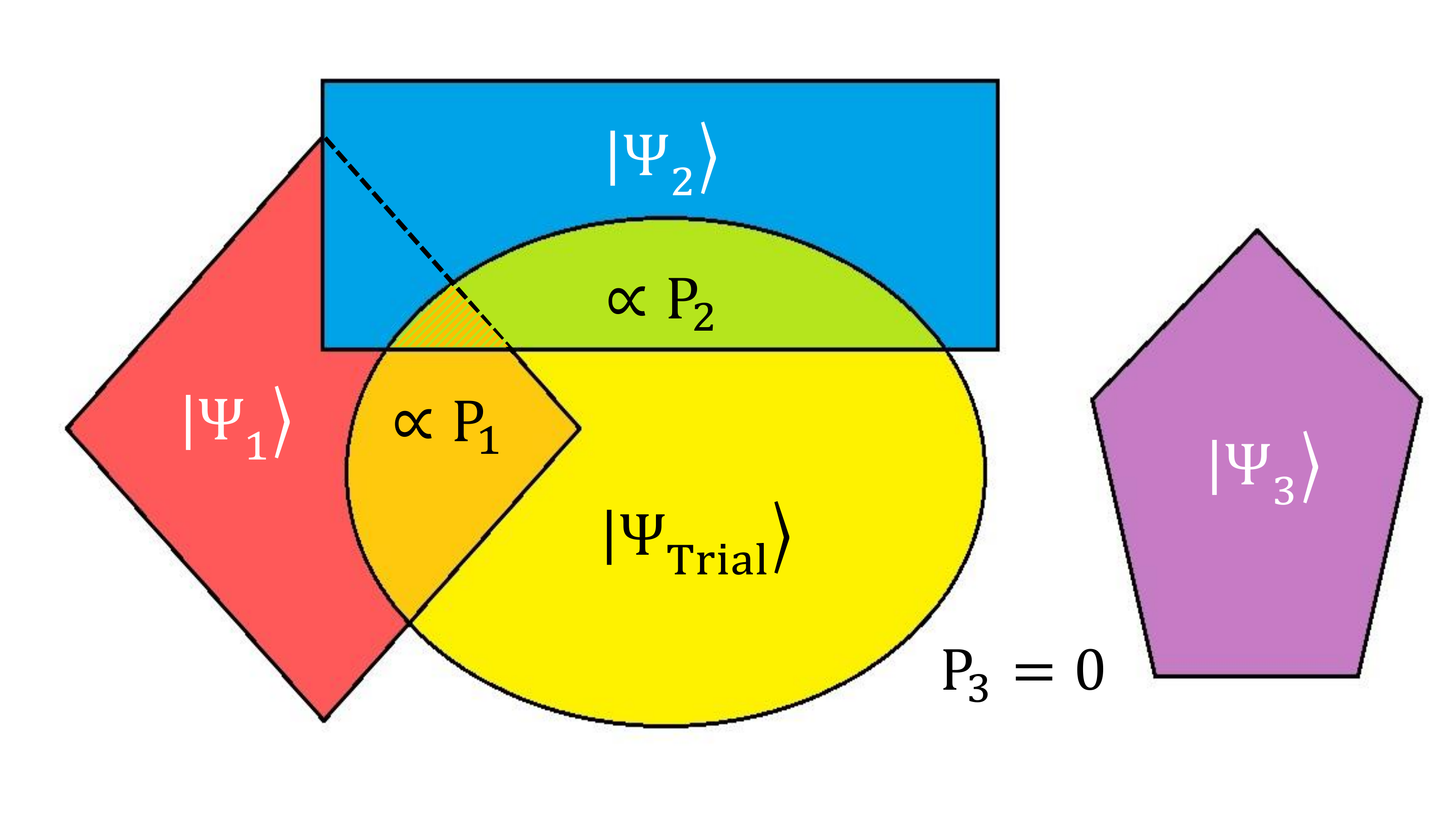}
\caption{An abstract representation of how a trial wavefunction may have overlap
with several states ($\ket{\Psi_{1}}$ and $\ket{\Psi_{2}}$) and be orthogonal
to others ($\ket{\Psi_{3}}$). In this figure, P$_i$ represents the probability of 
sampling the $i$-th state in a simulation with the QPE algorithm.}
\label{TrialWFFig}
\end{figure}

As described in Ref.~\cite{low2019q}, the distribution of energies for the ground and excited states 
from the QPE algorithm is determined by the Hamiltonian and a trial wavefunction composed of a superposition of 
Slater determinants. As abstractly illustrated in~\cref{TrialWFFig}, the trial wavefunction may have overlap
with several states ($\ket{\Psi_{1}}$ and $\ket{\Psi_{2}}$ in~\cref{TrialWFFig} and be orthogonal
to others ($\ket{\Psi_{3}}$ in~\cref{TrialWFFig}. The probability of obtaining an energy estimate for a 
particular state is proportional to the amount of overlap of the trial wave with that state's corresponding wave 
function, relative to all other overlaps. Through repeated simulations, one can obtain a distribution of energies 
for several states at the same time. The stochastic nature of the QPE algorithm is unlike any other current 
quantum algorithm which only provide energy estimates for a single targeted state. This also opens up opportunities
to find and chronicle exotic and novel states that are unobtainable with conventional computing and current 
approximate formulations.

 \begin{figure}
\includegraphics[clip,angle=0, width=0.45\textwidth]{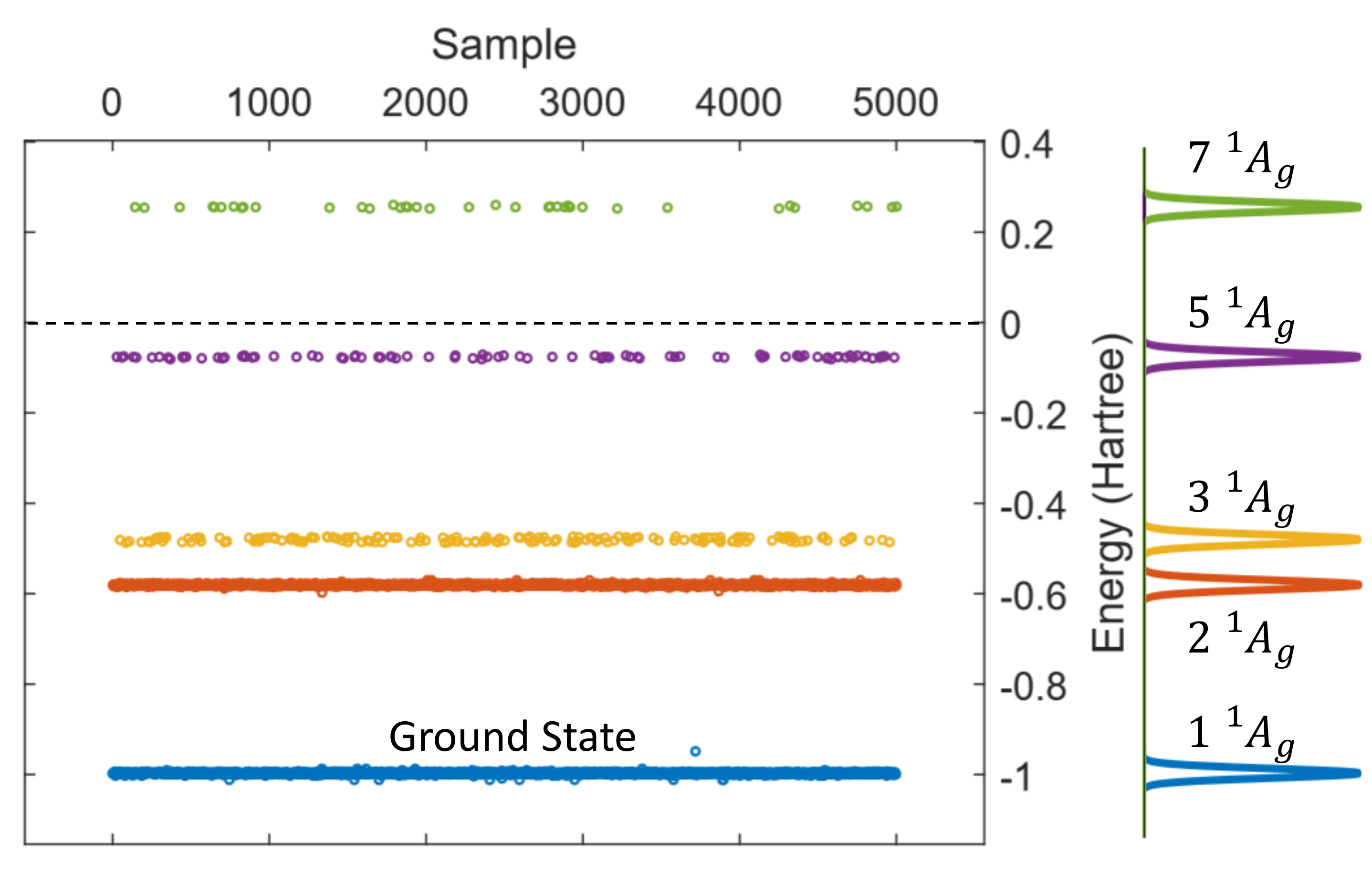}
\caption{A typical distribution of energies for a strongly correlated
system obtained from several simulations with the QPE algorithm. This particular 
spread of energies corresponds to the DUCC-ex($2\prescript{1}{}{A}_g$) results for H$_2$
at a stretched geometry ($R_{\rm H-H}=10$ a.u.) (see~\cref{Results}).}
\label{HistogramFig}
\end{figure}

\section{Numerical tests}\label{Results}

\begin{figure}
\includegraphics[trim={1.5cm 0 4.0cm 0},clip,angle=0, width=0.45\textwidth]{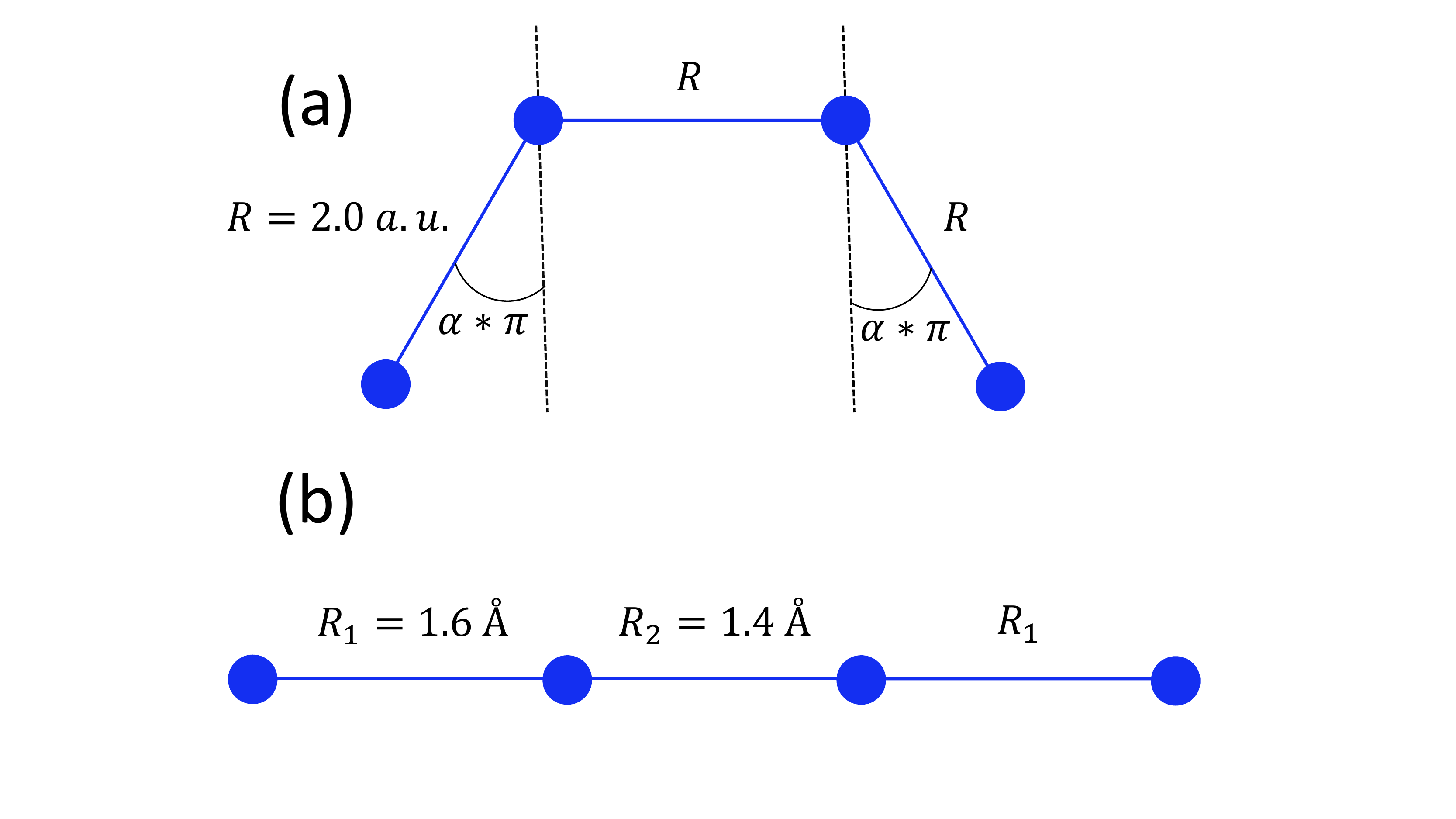}
\caption{H$_4$ models employed in excited-state simulations (see text for details).}
\label{h4}
\end{figure}

We  performed a series of numerical tests with the QDK simulator~\cite{low2019q}
for three systems characterized by strong ground-state correlation effects and 
low-lying singlet excited states defined by complicated multi-reference configurational 
structures: (1) H$_2$ systems for H--H separation corresponding to the equilibrium 
($R_{\rm H-H}=1.4008$ a.u.) and a stretched geometry ($R_{\rm H-H}=10$ a.u.), 
(2) H$_4$ system in the trapezoidal configuration corresponding to geometrical parameter 
$\alpha$ equal to 0.001 (see~\cref{h4}(a)) - for this geometry one can observe a 
strong quasi-degeneracy of low-lying electronic states, and (3) linear form of H$_4$ used 
in studies of singlet fission processes (see~\cref{h4}(b)). Special attention is paid to 
singlet doubly excited states which pose a significant challenge to existing many-body 
methodologies. For systems considered here we employ the cc-pVTZ basis set~\cite{dunning89_1007} 
and we correlate all electrons in all calculations. Since the QDK cannot exploit spatial 
symmetry, all QPE simulations were performed using $C_1$ symmetry. For simplicity, 
we will refer to the $D_{2h}$, $C_{2v}$, and $D_{2h}$ irreducible representations of the H$_2$, 
H$_4$ trapezoidal, and H$_4$ linear systems, respectively, when reporting full configuration 
interaction ((FCI) numbers obtained with  NWChem \cite{valiev10_1477}
(or equivalently, EOMCCSD or EOMCCSDTQ 
results for the H$_2$ and H$_4$ models, respectively). 
%

For H$_2$ models, we examine the performance of the bare and DUCC-ex transformed Hamiltonians 
for excited states, which have a significant (A) or partial (PA) amount of leading 
characteristic excitations in the active space, which contains one occupied and three lowest-lying 
RHF orbitals. Energies of these states for $R_{\rm H-H}=1.4008$ a.u. and $R_{\rm H-H}=10.0$ a.u. 
are shown in~\cref{table_h2eq,table_h2}, respectively. 
In~\cref{table_h2eq}, we consider three states $2\prescript{1}{}{A}_g$, $3\prescript{1}{}{A}_g$, and 
$5\prescript{1}{}{A}_g$, which are either comprehensively described within the active space ($2\prescript{1}{}{A}_g$) or 
have a significant overlap with it ($3\prescript{1}{}{A}_g$ and $5\prescript{1}{}{A}_g$ states). The $2\prescript{1}{}{A}_g$ is dominated by 
$\ket{\Phi_1^3}$ and $\ket{\Phi_{\bar{1}}^{\bar{3}}}$ single excitations, while the two other 
states have non-negligible out-of-active-space components. For example, $3\prescript{1}{}{A}_g$ state is dominated by
$\ket{\Phi_{1\bar{1}}^{2\bar{2}}}$ excitation but contains important component originating in the
$\ket{\Phi_{1}^{7}}$ and $\ket{\Phi_{\bar{1}}^{\bar{7}}}$ excitations. In a similar way, the 
$5\prescript{1}{}{A}_g$ state is dominated by $\ket{\Phi_{1\bar{1}}^{3\bar{3}}}$ excitations but it also contains 
important contributions from $\ket{\Phi_{1\bar{1}}^{3\bar{7}}}$ and $\ket{\Phi_{1\bar{1}}^{7\bar{3}}}$ 
configurations, which do not belong to the active space.
\renewcommand{\tabcolsep}{0.1cm}
\begin{center}
	\begin{table*}[ht]
		\centering
		\caption{Total energies of low-lying singlet excited states of H$_2$ system ($R_{\rm H-H}=1.4008$ a.u.) of the $A_g$ symmetry (all classical calculations for H$_2$ have been performed using $D_{2h}$ symmetry group). The values in parenthesis are errors relative to FCI. In all calculations, restricted Hartree-Fock orbitals were used.}
\begin{tabular}{lccccc} \hline  \\	
\multicolumn{1}{l}{State }	& \multicolumn{1}{c}{FCI$^{(a)}$} & 
\multicolumn{1}{c}{DUCC-ex($2\prescript{1}{}{A}_g$)   }  &
\multicolumn{1}{c}{DUCC-ex($3\prescript{1}{}{A}_g$)   }  &
\multicolumn{1}{c}{DUCC-ex($5\prescript{1}{}{A}_g$)   }  &	
\multicolumn{1}{c}{bare Hamiltonian }   \\[-0.0cm]
\multicolumn{1}{c}{(Char.)$^{(b)}$} &    &
\multicolumn{1}{c}{ini. SD: $\frac{1}{\sqrt{2}}
(\ket{\Phi_1^3} + \ket{\Phi_{\bar 1}^{\bar 3}})$}	 & 
\multicolumn{1}{c}{ini. SD: $\ket{\Phi_{1\bar{1}}^{2\bar{2}}}$}   & 
\multicolumn{1}{c}{ini. SD: $\ket{\Phi_{1\bar{1}}^{3\bar{3}}}$}   &  
\multicolumn{1}{c}{ini. SD: $\ket{\Phi_{1\bar{1}}^{2\bar{2}}}$}
\\[0.1cm]
			\hline\hline \\  
%
$2\prescript{1}{}{A}_g$ & -0.5487  & -0.5478 $\pm$ 0.0066 & -0.5197 $\pm$ 0.0019 & -0.5034 $\pm$ 0.0030 & -0.5306 $\pm$ 0.0017 \\[-0.0cm]
    (A)  &          &  (0.0009)           & (0.0290)            & (0.0453)            & (0.0181)           \\[0.3cm]
%
$3\prescript{1}{}{A}_g$ & -0.1210  & ----- & 0.0644 $\pm$ 0.0017 & -0.0696 $\pm$ 0.0016 & -0.0622 $\pm$ 0.0018  \\[-0.0cm]
  (PA)   &          &       & (0.0566)          & (0.0514)           & (0.0588)            \\[0.3cm]
%
$5\prescript{1}{}{A}_g$ &   0.2310 & 0.2914 $\pm$ 0.0021 & 0.2819 $\pm$ 0.0065 & 0.2671 $\pm$ 0.0031 & 0.2842 $\pm$ 0.0014 \\[-0.0cm]
    (PA) &          & (0.0604)          & (0.0508)          & (0.0361)          & (0.0532)         \\[0.3cm]
\hline\end{tabular}
\label{table_h2eq}
\footnotetext[0]{
\setlength{\baselineskip}{1em} 
{$^{(a)}$ Full configuration interaction calculations were performed using all 30 molecular orbitals.    $^{(b)}$ Character of the electronic states is determined by the active-space contribution: A - dominated by active-space configurations, PA -  dominated by configurations not belonging to active space. }
}
\end{table*}
\end{center}

When the bare Hamiltonian for H$_2$ at equilibrium is diagonalized in the active space, 
all three excited states can be observed. To improve the distribution and get a better
statistical sampling for the excited states (particularly $3\prescript{1}{}{A}_g$), the initial wavefunction was
chosen to be the $\ket{\Phi_{1\bar{1}}^{2\bar{2}}}$ configuration, although it is not necessary to 
track these excited states. Unfortunately, there is a significant difference between the FCI 
energies and energies obtained by diagonalizing the bare Hamiltonian with errors of over 
18 milliHartree for $2\prescript{1}{}{A}_g$ and 50 milliHartree for $3\prescript{1}{}{A}_g$ and $5\prescript{1}{}{A}_g$. 

The DUCC-ex formalism for the $2\prescript{1}{}{A}_g$ state (DUCC-ex($2\prescript{1}{}{A}_g$)), where the similarity transformation 
is driven by $\sigma_{\rm ext}(K)$ corresponding to the $2\prescript{1}{}{A}_g$ state (see~\cref{s1ext,s2ext}), provides an excellent agreement to the FCI energies with an error less than 1 milliHartree, 
a substantial improvement over the energy obtained with the bare Hamiltonian. From~\cref{table_h2eq}, 
one can notice that by using the trial wavefunction $\frac{1}{\sqrt{2}}(\ket{\Phi_1^3} + 
\ket{\Phi_{\bar 1}^{\bar 3}})$, there is a chance to describe not only target $2\prescript{1}{}{A}_g$ (which given 
the state-specific nature of the DUCC-ex approach is the physical solution) but also other states, 
which have non-zero overlap with the trial wavefunction (in this case $5\prescript{1}{}{A}_g$). For the $3\prescript{1}{}{A}_g$ and
$5\prescript{1}{}{A}_g$ states, the effect of important out-of-active-space Slater determinants needs to be determined perturbatively and is only approximately captured by the corresponding $\sigma_{\rm ext}(K)$ operators.
Consequently, the DUCC-ex($3\prescript{1}{}{A}_g$) and DUCC-ex($5\prescript{1}{}{A}_g$) results are less accurate compared to the 
DUCC-ex($2\prescript{1}{}{A}_g$) case. 
However, the results obtained with the respective DUCC-ex($3\prescript{1}{}{A}_g$) and DUCC-ex($5\prescript{1}{}{A}_g$) Hamiltonians 
are still better compared to the diagonalization of bare-Hamiltonian. Also, in the simulations for $3\prescript{1}{}{A}_g$ and $5\prescript{1}{}{A}_g$ all 
three states are observed when the initial wavefunctions are taken to be the leading excitation for the corresponding state.

Compared to the equilibrium geometry,  the H$_2$ system corresponding to the $R_{\rm H-H}=10.0$ a.u. is characterized by
a larger number of excited states with a significant overlap
with active space. For $R_{\rm H-H}=10.0$ a.u. one can identify five excited states of $A_g$ symmetry falling into this category. 
As seen from~\cref{table_h2} the low-lying excited states can be efficiently described by the DUCC-ex formalism. 
For example, the DUCC-ex($2\prescript{1}{}{A}_g$) and DUCC-ex($3\prescript{1}{}{A}_g$) Hamiltonians provide very accurate estimates of the $2\prescript{1}{}{A}_g$ and 
$3\prescript{1}{}{A}_g$ states, which is quite remarkable given the size of the active space. In both cases, one can observe 
a significant improvement of the results obtained by the diagonalization of the bare Hamiltonian in the same 
active space, which provides a good illustration of the efficiency of the DUCC-ex formalism, even in a simple 
case corresponding to single commutator expansion of the DUCC-ex downfolded Hamiltonian. For higher excited 
states ($5\prescript{1}{}{A}_g$, $6\prescript{1}{}{A}_g$, $7\prescript{1}{}{A}_g$), the DUCC-ex results are less accurate, but still comparable to the 
results obtained through the diagonalization of bare Hamiltonian in the active space. This behavior is 
justified given the complex excitation manifold describing the higher-lying states and ought to be resolved 
with improved approximations for $\sigma_{\rm ext}(K)$.

\renewcommand{\tabcolsep}{0.05cm}
\begin{center}
\begin{table*}
\centering
\caption{Total energies of low-lying singlet excited states of H$_2$ system ($R_{\rm H-H}=10$ a.u.) of the $A_g$ symmetry (all classical calculations for H$_2$ have been performed using $D_{2h}$ symmetry group). The values in parenthesis are errors relative to FCI. In all calculations, restricted Hartree-Fock orbitals were used.}
\begin{tabular}{lcccccc} \hline  \\
\multicolumn{1}{l}{State }	& \multicolumn{1}{c}{FCI$^{(a)}$} & 
\multicolumn{1}{c}{DUCC-ex($2\prescript{1}{}{A}_g$)}  &
\multicolumn{1}{c}{DUCC-ex($3\prescript{1}{}{A}_g$)}  &
\multicolumn{1}{c}{DUCC-ex($5\prescript{1}{}{A}_g$)}  &	
\multicolumn{1}{c}{DUCC-ex($7\prescript{1}{}{A}_g$)   }  &	
\multicolumn{1}{c}{bare Hamiltonian}  \\[-0.0cm]
(Char.)$^{(b)}$ & 	&	
\multicolumn{1}{c}{ini. SD: $\ket{\Phi_{1\bar{1}}^{2\bar{2}}}$}	 & 
\multicolumn{1}{c}{ini. SD: $\frac{1}{\sqrt{2}}
(\ket{\Phi_1^3} + |\Phi_{\bar 1}^{\bar 3})$}   & 
\multicolumn{1}{c}{ini. SD: $\ket{\Phi_{1\bar{1}}^{3\bar{3}}}$}   & 
\multicolumn{1}{c}{ini. SD: $\ket{\Phi_{1\bar{1}}^{4\bar{4}}}$}
&
\multicolumn{1}{c}{ini. SD: $\ket{\Phi_{1\bar{1}}^{2\bar{2}}}$}  
\\[0.1cm]
\hline\hline \\  
%
%
$2\prescript{1}{}{A}_g$ & -0.5981 & -0.6047 $\pm$ 0.0018 & -0.5847 $\pm$ 0.0018 & -0.5517 $\pm$ 0.0018 & -0.5610 $\pm$ 0.0013 & -0.5803 $\pm$ 0.0019 \\[-0.0cm]
   (A)   &         & (-0.0065)          & (0.0134)           & (0.0464)           & (0.0371)           & (0.0178)    \\[0.3cm]
%
$3\prescript{1}{}{A}_g$ & -0.4873 & -0.4825 $\pm$ 0.0045 & -0.4848 $\pm$ 0.0050 & -0.4747 $\pm$ 0.0043 & -0.4785 $\pm$ 0.0035 & -0.4795 $\pm$ 0.0043 \\[-0.0cm]
  (A)    &         & (0.0048)           & (0.0025)           & (0.0126)           & (0.0088)           & (0.0078)  \\[0.3cm]
%
%
$5\prescript{1}{}{A}_g$ & -0.1040 & -0.0812 $\pm$ 0.0018 & -0.0779 $\pm$ 0.0017 & -0.0625 $\pm$ 0.0018 & -0.0592 $\pm$ 0.0020 & -0.0763 $\pm$ 0.0017 \\[-0.0cm]
  (A)    &         & (0.0228)           & (0.0261)           & (0.0415)           & (0.0448)           & (0.0277) \\[0.3cm]
%
$6\prescript{1}{}{A}_g$ &  0.0238 & ----- & ------ &  0.0405 $\pm$ 0.0016 & 0.0365 $\pm$ 0.0017 & 0.0368 $\pm$ 0.0024    \\[-0.0cm]
 (A)     &         &       &        & (0.0167)           & (0.0127)          & (0.0130) \\[0.3cm]
%
$7\prescript{1}{}{A}_g$ &  0.2326 & 0.2525 $\pm$ 0.0023 & 0.2545 $\pm$ 0.0019 & 0.2623 $\pm$ 0.0019 & 0.2584 $\pm$ 0.0017 & 0.2556 $\pm$ 0.0020 \\[-0.0cm]
 (A)     &         & (0.0199)          & (0.0219)          & (0.0297)          & (0.0258)          & (0.0230)  \\[0.3cm]
%
%
\hline\end{tabular}
\label{table_h2}
\footnotetext[0]{
\setlength{\baselineskip}{1em} 
{$^{(a)}$ Full configuration interaction calculations were performed using all 30 molecular orbitals. $^{(b)}$ Character of the electronic states is determined by the active-space contribution: A - dominated by active-space configurations, PA -  dominated by configurations not belonging to active space.}
}
\end{table*}
\end{center}

\cref{table_h2} also provides an excellent illustration of the fact that trial wavefunction can be used to probe various electronic states in quantum simulations. For example, using the trial state $\ket{\Phi_{1\bar{1}}^{2\bar{2}}}$ for QPE simulations of the DUCC-ex($2\prescript{1}{}{A}_g$) Hamiltonian one can obtain a statistically meaningful presence of states different from challenging doubly-excited $2\prescript{1}{}{A}_g$ state, i.e., $1\prescript{1}{}{A}_g$ (not shown in the~\cref{table_h2}), $3\prescript{1}{}{A}_g$, $5\prescript{1}{}{A}_g$, and $7\prescript{1}{}{A}_g$. 
This fact is associated with the strong quasi-degeneracy of the ground-state and low-lying excited states with contribution from the  $\ket{\Phi_{1\bar{1}}^{2\bar{2}}}$ Slater determinant.
It is also worth mentioning that all  states considered here are  either purely doubly excited states or states of mixed single- and doubly-excited character, which usually pose a significant challenge for approximate EOMCC formulations. 
Analogous  situations are naturally occurring in strongly correlated systems and the fact that QPE can deal with these challenging problems provides yet another argument in favor of developing quantum algorithms for excited states. 
Similar behavior can also be seen in the case of the active-space representation of the bare Hamiltonian. 
Although the DUCC-ex approach is state-specific in its nature, the appearance  of other physically interpretable excited states in the spectra of 
DUCC-ex Hamiltonians should be explored further.

We exploited the state-specific nature of the DUCC-ex approach in an attempt to 
describe the lowest-lying fully symmetric $2\prescript{1}{}{A}_1$ and $2\prescript{1}{}{A}_g$ states ({\it vide infra}) 
of the H$_4$ system in trapezoidal ($\alpha=0.001$) (H$_4$(a) system) and linear 
configuration (H$ _4$(b) system) shown in~\cref{h4}. For both systems, FCI results 
were obtained by running EOMCCSDTQ calculations using $C_{2v}$ and $D_{2h}$ symmetries, 
respectively. The active space used to construct DUCC effective Hamiltonians in both configurations includes the two occupied
orbitals and the five lowest-lying virtual orbitals. While the ground state for H$_4$(a) discloses strong quasidegeneracy 
effects between $\ket{\Phi}$ and $\ket{\Phi_{2\bar{2}}^{3\bar{3}}}$ Slater determinants,
for the H$_4$(b) system these effects slightly weaker. 
In the description of $2\prescript{1}{}{A}_1$ state of the H$_4$(a) system the dominant role is played by the
$\ket{\Phi_{2\bar{2}}^{3\bar{3}}}$ determinant, which almost entirely dictates
the corresponding wavefunction expansion.
The  $2\prescript{1}{}{A}_g$ state of H$_4$(b) reveals a more multi-configurational structure where determinants
$\ket{\Phi_1^3}$, $\ket{\Phi_{\bar 1}^{\bar 3}}$, $\ket{\Phi_2^4}$, $\ket{\Phi_{\bar 2}^{\bar 4}}$,
$\ket{\Phi_{12}^{34}}$, $\ket{\Phi_{\bar{1}\bar{2}}^{\bar{3}\bar{4}}}$,
$\ket{\Phi_{1\bar{1}}^{3\bar{3}}}$, $\ket{\Phi_{2\bar{2}}^{3\bar{3}}}$, and
$\ket{\Phi_{2\bar{2}}^{4\bar{4}}}$ all have a non-negligible contribution to the wavefunction. 
As seen from~\cref{table_h4}, we are able to inspect the lowest-lying doubly excited state
for both systems with the bare Hamiltonian. It is important to note, that unlike the H$_2$ systems, 
one must use an initial guess that contains the $\ket{\Phi_{2\bar{2}}^{3\bar{3}}}$ determinant with some significant weight. Otherwise, only the ground state will be observed. Even though these states can be
tracked with the bare Hamiltonians, they come with errors of 73 millihartree for the H$_4$(a) system and 
38 millihartree for the H$_4$(b) system. As shown in~\cref{table_h4}, the DUCC-ex($2\prescript{1}{}{A}_1$) approach 
can very efficiently reproduce the FCI $2\prescript{1}{}{A}_1$ energy with relative error less than 10 milliHartree. 
As in the case of H$_2$ system using $\ket{\Phi_{2\bar{2}}^{3\bar{3}}}$ as a initial state for quantum 
simulations one can observe a statistical presence of states which have a non-negligible overlap with this 
determinant. It is interesting to notice that for the H$_4$(b) system, the DUCC-ex results for the $2\prescript{1}{}{A}_g$ 
are less accurate and they provide comparable results to the bare Hamiltonian energy. In our opinion, this 
is associated with the strong multi-configurational character of the corresponding wavefunction, where 
simple approximations~\cref{s1exp,s2exp} may not provide the desired level of accuracy.

\renewcommand{\tabcolsep}{0.4cm}
\begin{center}
\begin{table*}
\centering
\caption{Total energies of low-lying singlet states of the H$_4$ system of the $A_1$ (H$_4$(a)) or $A_g$ (H$_4$(b)) symmetry. The classical calculations for H$_4$ have been performed using either the $C_{2v}$ (H$_4$(a)) or $D_{2h}$ (H$_4$(b)) symmetry group. The values in parenthesis are errors relative to FCI. In all calculations, restricted Hartree-Fock orbitals were used.}
\begin{tabular}{lccc} \hline  \\
\multicolumn{4}{c}{$\alpha$=0.001 (H$_4$(a) system)} \\[0.2cm]
\multicolumn{1}{l}{State }	& \multicolumn{1}{c}{FCI$^{(a)}$} & bare Hamiltonian &
\multicolumn{1}{c}{DUCC-ex($2\prescript{1}{}{A}_1$)} \\[0.0cm]
(Char.)$^{(b)}$ &  & \multicolumn{1}{c}{ini. SD: $\ket{\Phi_{2\bar{2}}^{3\bar{3}}}$}	& \multicolumn{1}{c}{ini. SD: $\ket{\Phi_{2\bar{2}}^{3\bar{3}}}$} \\[0.1cm]
\hline\hline \\
%
$2\prescript{1}{}{A}_1$ & -2.0280  & -1.9546 $\pm$ 0.0037 & -2.0184 $\pm$ 0.0033 \\[-0.0cm]
 (A)     &          & (0.0734) & (0.0096) 
\\[0.2cm] \\
\multicolumn{4}{c}{linear (H$_4$(b) system)} \\[0.2cm]
\multicolumn{1}{l}{State }	& \multicolumn{1}{c}{FCI$^{(a)}$} & bare Hamiltonian &
\multicolumn{1}{c}{DUCC-ex($2\prescript{1}{}{A}_g$)} \\[0.0cm]
(Char.)$^{(b)}$ &  & \multicolumn{1}{c}{ini. SD: $\ket{\Phi_{2\bar{2}}^{3\bar{3}}}$}	& \multicolumn{1}{c}{ini. SD: $\ket{\Phi_{2\bar{2}}^{3\bar{3}}}$} \\[0.1cm]
\hline\hline \\
%
$2\prescript{1}{}{A}_g$ & -1.9901 & -1.9513 $\pm$ 0.0052 & -1.9458 $\pm$ 0.002894  \\[-0.0cm]
   (A)   &         & (0.0388)           & (0.0443)  \\[0.3cm]
\hline\end{tabular}
\label{table_h4}
\footnotetext[0]{
\setlength{\baselineskip}{1em} 
{$^{(a)}$ Full configuration interaction calculations were performed using all 60 molecular orbitals. $^{(b)}$ Character of the electronic states is determined by the active-space contribution: A - dominated by active-space configurations, PA -  dominated by configurations not belonging to active space.}
}
\end{table*}
\end{center}

\section{Conclusions}
In this paper, we discussed the excited-state extension of the DUCC formalism and its amenability for quantum computing. Using simple approximation schemes based on the utilization of the EOMCCSD wavefunction representation and lowest-rank contribution stemming from a single commutator electronic Hamiltonian and external $\sigma_{\rm ext}(K)$ operator, we were able to demonstrate that the active-space representation of  downfolded Hamiltonians can be used to reproduce a large portion of excited-state correlation effects. For H$_2$ and H$_4$(a) models, we observed significant improvements in excited-state energies compared to the diagonalization of bare Hamiltonian in the active space, which was especially true for low-lying states dominated by active-space contributions (usually attributed to singly excited states and low-lying doubly excited states).
As expected, for active-space dominated states characterized by  higher excitation energies and more complicated configurational structure, the efficiency of a simple approximation schemes discussed here deteriorates, which is indicative of the need for the inclusion of  higher-order commutators and more efficient estimates of the $\sigma_{\rm ext}(K)$ operators. This issue  will be explored in future research  by coupling DUCC-ex formalism with higher-rank EOMCC formulations. 
We have also demonstrated that quantum phase estimation can provide an efficient tool for testing various "excited-state" hypothesis for strongly correlated systems, which usually pose a significant challenge for existing many-body formalism due to the  need of inclusion of higher rank excitation effects and high density of states in a narrow energy gap, which renders numerical  identification of state of interest  numerically unfeasible. 
Instead, using QPE techniques one can define hypothesis or trial state and obtains in the course of calculation statistical footprint of all states that have non-negligible overlap with the hypothesis state. Although this fact is a well-known foundation of quantum computing, it deserves broader exposure. The stochastic character of QPE may be very useful in studies of excited-state processes, especially in strongly correlated or metallic-like system.


\section{acknowledgement}

This  work  was  supported  by  the "Embedding Quantum Computing into Many-body Frameworks for Strongly Correlated  Molecular and Materials Systems" project, 
which is funded by the U.S. Department of Energy(DOE), Office of Science, Office of Basic Energy Sciences, the Division of Chemical Sciences, Geosciences, and Biosciences.
A portion of this research was funded by the  Quantum Algorithms, Software, and Architectures (QUASAR) Initiative at Pacific Northwest National Laboratory (PNNL). It was conducted under the Laboratory Directed Research and Development Program at PNNL.
All calculations have been performed using the Molecular Science Computing Facility (MSCF) in the Environmental Molecular Sciences Laboratory (EMSL) at the Pacific Northwest National Laboratory (PNNL). EMSL is funded by the Office of Biological and Environmental Research in the U.S. Department of Energy. PNNL is operated for the U.S. Department of Energy by the Battelle Memorial Institute under Contract DE-AC06-76RLO-1830.

\end{document}